\documentclass[aps,pra,twocolumn,floatfix,superscriptaddress,footinbib]{revtex4-1}

\usepackage[utf8]{inputenc}
\usepackage[english]{babel}
\usepackage[T1]{fontenc}
\usepackage{amsmath, amsfonts, amssymb, amsthm, mathrsfs, amssymb, braket, bbm, bm,graphicx,times,txfonts,color,subfigure}
\usepackage{hyperref}

\usepackage{multirow}

\begin{document}

\title{There is more to quantum interferometry than entanglement}

\author{Thomas~R.~Bromley}
\email{thomas.r.bromley@gmail.com}
\affiliation{Centre for the Mathematics and Theoretical Physics of Quantum Non-Equilibrium Systems, School of Mathematical Sciences, The University of Nottingham, University Park, Nottingham NG7 2RD, UK}

\author{Isabela~A.~Silva}
\affiliation{Centre for the Mathematics and Theoretical Physics of Quantum Non-Equilibrium Systems, School of Mathematical Sciences, The University of Nottingham, University Park, Nottingham NG7 2RD, UK}
\affiliation{Instituto de F{\'{i}}sica de S{\~{a}}o Carlos, Universidade de S{\~{a}}o Paulo, P.O. Box 369, S{\~{a}}o Carlos, 13560-970 S{\~{a}}o Paulo, Brazil}

\author{Charlie~O.~Oncebay-Segura}
\affiliation{Instituto de F{\'{i}}sica de S{\~{a}}o Carlos, Universidade de S{\~{a}}o Paulo, P.O. Box 369, S{\~{a}}o Carlos, 13560-970 S{\~{a}}o Paulo, Brazil}

\author{Diogo~O.~Soares-Pinto}
\affiliation{Instituto de F{\'{i}}sica de S{\~{a}}o Carlos, Universidade de S{\~{a}}o Paulo, P.O. Box 369, S{\~{a}}o Carlos, 13560-970 S{\~{a}}o Paulo, Brazil}

\author{Eduardo~R.~deAzevedo}
\affiliation{Instituto de F{\'{i}}sica de S{\~{a}}o Carlos, Universidade de S{\~{a}}o Paulo, P.O. Box 369, S{\~{a}}o Carlos, 13560-970 S{\~{a}}o Paulo, Brazil}

\author{Tommaso~Tufarelli}
\affiliation{Centre for the Mathematics and Theoretical Physics of Quantum Non-Equilibrium Systems, School of Mathematical Sciences, The University of Nottingham, University Park, Nottingham NG7 2RD, UK}

\author{Gerardo~Adesso}
\affiliation{Centre for the Mathematics and Theoretical Physics of Quantum Non-Equilibrium Systems, School of Mathematical Sciences, The University of Nottingham, University Park, Nottingham NG7 2RD, UK}

\begin{abstract}
 Entanglement has long stood as one of the characteristic features of quantum mechanics, yet recent developments have emphasized the importance of quantumness beyond entanglement for quantum foundations and technologies. We demonstrate that entanglement cannot entirely capture the worst-case sensitivity in quantum interferometry, when quantum probes are used to estimate the phase imprinted by a Hamiltonian, with  fixed energy levels but variable eigenbasis,  acting on one arm of an interferometer. This is shown by defining a bipartite entanglement monotone tailored to this interferometric setting and proving that it never exceeds the so-called interferometric power, a quantity which relies on more general quantum correlations beyond entanglement and captures the relevant resource.  We then prove that the interferometric power can never increase when local commutativity-preserving operations are applied to qubit probes, an important step to validate such a quantity as a genuine quantum correlations monotone. These findings are accompanied by a room-temperature nuclear magnetic resonance experimental investigation, in which two-qubit states with extremal (maximal and minimal) interferometric power at fixed entanglement are produced and characterized.
\end{abstract}


\date{\today}
\maketitle


\section{Introduction}

Entanglement is often perceived as the beating heart of quantum technologies~\cite{nielsen2010quantum,o2009photonic,horodecki2009quantum}. It is the power behind a wealth of processes crucial for current innovations~\cite{vedral2014quantum,dowling2003quantum}, including quantum computing~\cite{steane1998quantum,bennett1997strengths} and cryptography~\cite{gisin2002quantum}. Moreover, comprehending entanglement is of fundamental importance for quantum foundations~\cite{horodecki2009quantum}, helping to demarcate the ever elusive boundary between classical physics and truly quantum phenomena~\cite{haroche2006exploring,haroche2013nobel}. It is clear that what Schr{\"{o}}dinger once termed \emph{the} characteristic trait of quantum mechanics is central to our quantum journey~\cite{schrodinger1935discussion}. However, the seemingly indisputable role of entanglement has been recently challenged by the idea of quantum correlations \emph{beyond} entanglement~ \cite{ollivier2001quantum,henderson2001classical,modi2012classical,adesso2016measures,adesso2016intro}. These more general correlations, accounting for the inevitable disturbance caused by a local measurement on one subsystem of a genuinely quantum state, can play their own part in quantum enhanced processes, ranging from entanglement distribution to quantum state merging~\cite{adesso2016measures,chuan2012quantum,streltsov2012quantum,cavalcanti2011operational,madhok2011interpreting}.

Metrology, the science of high precision measurement, is one of the quintessential fields experiencing an advantage in the presence of entanglement: it has long been appreciated that measurements can be performed to greater precision by using an entangled collection of probes~\cite{giovannetti2006quantum,giovannetti2011advances,toth2014quantum,pezze2016nonclassical}. Despite this, the exact role of entanglement in the related task of quantum interferometry, which has far-reaching applications such as gravitational wave detection \cite{abbott2016observation}, remains unclear \cite{braun2017quantum}. The goal here is to precisely measure a phase shift $\varphi$ by passing two quantum probes in a bipartite state $\rho_{AB}$ through different arms of an interferometer~\cite{caves1981quantum,demkowicz2015quantum}. One arm actively imprints the phase $\varphi$ onto probe $A$ by a unitary transformation $U_{A}^{\varphi}=e^{- i \varphi H_{A}}$, generated by the Hamiltonian $H_{A}$, while the other arm leaves probe $B$ unchanged. An estimate $\tilde{\varphi}$ of the parameter is then constructed by carrying out suitable measurements on $\nu$ copies of the output state of the probes $\rho_{AB}^{\varphi} = (U_{A}^{\varphi}\otimes \mathbb{I}_{B}) \rho_{AB}(U_{A}^{\varphi}\otimes \mathbb{I}_{B})^{\dagger}$.

This estimate has an associated precision given by the mean squared error $\Delta^{2} \tilde{\varphi}$, quantifying the statistical distance between $\tilde{\varphi}$ and $\varphi$. The objective is to reach the highest precision possible, but this is always limited by the Cram{\'{e}}r-Rao bound $\Delta^{2} \tilde{\varphi} \geq [\nu \mathcal{F}(\rho_{AB},H_{A})]^{-1}$~\cite{helstrom1976quantum}, with $\mathcal{F}(\rho_{AB},H_{A})$ being the quantum Fisher information (QFI)~\cite{braunstein1994statistical,paris2009quantum}. In the asymptotic limit $\nu \rightarrow \infty$, this bound can be saturated if optimal measurements are performed on the probes. The QFI thus stands as the relevant figure of merit in interferometry, capturing the sensitivity of $\rho_{AB}$ to phase imprinting with a known Hamiltonian $H_{A}$.

A recent series of works \cite{girolami2013characterizing,girolami2014quantum,adesso2014gaussian,farace2016building,nichols2016practical} investigated the scenario where only the energy-level spectrum $\Gamma$ (assumed to be non-degenerate) of the Hamiltonian $H_{A}$ is fixed \emph{a priori}, while its eigenbasis is  not known at the initial stage of preparation of the probe state, due, e.g.,~to environmental fluctuations or set rules of a game. The family of possible Hamiltonians used to imprint the phase $\varphi$ is then $H_{A}^{\Gamma} = V \,\mbox{diag}\left(\Gamma\right) V^{\dagger}$ for all unitaries $V$ on subsystem $A$. In this setting, one is interested in gauging the usefulness of a given probe state $\rho_{AB}$ for interferometry, regardless of the specific direction of phase imprinting.  Such an analysis may lead to the identification of versatile probe states that can be useful as resources for precise phase estimation in several different bases. In particular, it becomes important to assess the {\it worst-case sensitivity}, obtained when $H_{A}^{\Gamma}$ generates the nontrivial local dynamics (compliant with the fixed spectral constraint) to which the probe is the least sensitive. To test a probe state $\rho_{AB}$ in such unfavorable conditions, an adversarial referee is assumed to operate the Hamiltonian $H_A^{\Gamma}$ in a black box. The referee then reveals the selected eigenbasis only after the interaction, so that the most informative measurement given the prior preparation of $\rho_{AB}$ and this posterior information on $H_A^\Gamma$ can be performed on the output state $\rho_{AB}^\varphi$, in order to estimate $\varphi$. Then,  the relevant figure of merit for such a setup is the minimum QFI over all $H_A^\Gamma$, given by
\begin{equation}\label{Eq:IP}
\mathcal{P}_{A}^{\Gamma}(\rho_{AB})=\mbox{$\frac{1}{4}$}\min_{H_{A}^{\Gamma}}\mathcal{F}(\rho_{AB},H_{A}^{\Gamma}),
\end{equation}
including a convenient normalization factor. This quantity, which captures the worst-case sensitivity to phase imprinting  within the family of Hamiltonians $H_A^{\Gamma}$ achievable by a probe state $\rho_{AB}$, has been aptly baptized {\em interferometric power} (IP) in \cite{girolami2014quantum}. Interestingly, the interferometric power vanishes if and only if $\rho_{AB}$ is a {\it classical} state \cite{girolami2013characterizing,girolami2014quantum} of the form
\begin{equation}\label{Eq:CQ}
\rho_{AB} = {\sum}_{i}\, p_{i} \ket{i}\bra{i}_{A} \otimes \rho_{B}^{(i)},
\end{equation}
with $\{\ket{i}_{A}\}$ being any orthonormal basis of $A$ and $\rho_{B}^{(i)}$ being arbitrary states of $B$~\cite{adesso2016measures}. Notice that classical states of Eq.~(\ref{Eq:CQ}) form  a strict subset of {\it separable} states, which in turn  can be written as
\begin{equation}\label{Eq:Sep}
\rho_{AB} = {\sum}_{i}\, p_{i} \, \rho_{A}^{(i)} \otimes \rho_{B}^{(i)},
\end{equation}
with $\rho_{A}^{(i)}$ being arbitrary states of $A$. Therefore, the IP has been suggested as a quantifier of quantum correlations {\it beyond} entanglement in $\rho_{AB}$ (with respect to probe $A$) \cite{girolami2014quantum,adesso2016measures}. Operationally, a signature of these more general correlations is indeed the sensitivity to phase imprinting with all possible local Hamiltonian generators, that is, the ability to exhibit quantum {\it  coherence} \cite{streltsov2016colloquium} in all possible local bases for probe $A$ \cite{adesso2016measures}.


It is therefore natural to wonder where entanglement comes into play, if at all. In this paper, we shed light on this question in quantitative terms by showing that entanglement, once suitably quantified, accounts only for a partial contribution to the available precision in quantum interferometry, hence formalizing a hierarchy of quantum resources useful for this task. In Sec.~\ref{Sec:IE} we define a bipartite entanglement monotone, the {\em interferometric entanglement} (IE), specifically motivated from pure-state interferometry, and we show in Sec.~\ref{Sec:HI} that it never exceeds the IP for arbitrary mixed bipartite quantum states. The IP is further proven in Sec.~\ref{Sec:IP} to never increase when qubit probes are subjected to local commutativity-preserving operations~\cite{hu2012necessary,guo2013necessary,streltsov2011behavior}, which constitute a meaningful set of free operations for the sought-after resource theory of quantum correlations~\cite{adesso2016measures}. This realizes important progress towards establishing the IP as a fully fledged and operationally relevant quantum correlations monotone. In Sec.~\ref{Sec:EX} we then investigate the relationship between IP and entanglement experimentally with a room-temperature liquid-state nuclear magnetic resonance (NMR) implementation of a two-qubit system, in which case our IE reduces to the tangle (squared concurrence)~\cite{coffman2000distributed,hill1997entanglement,wootters1998entanglement,rungta2001universal,rungta2003concurrence} and the IP adopts a simple closed form~\cite{girolami2014quantum}. Rank-2 states with the largest and smallest IP for a fixed tangle are generated and characterized, demonstrating that highly mixed states containing extremal quantum correlations additional to entanglement are accessible in the laboratory, and could be adopted as robust probes for black box interferometry experiments \cite{girolami2014quantum}. We draw our concluding remarks in Sec.~\ref{Sec:SU}.





\section{Interferometric entanglement}\label{Sec:IE}
We define the IE for any pure bipartite probe state $\ket{\psi}_{AB}$ as
\begin{equation}\label{Eq:IEPure}
\mathcal{E}^{\Gamma}(\ket{\psi}_{AB})=\min_{H_{A}^{\Gamma}}{\cal V}(\ket{\psi}_{AB},H_{A}^{\Gamma}),
\end{equation}
with the variance ${\cal V}(\ket{\psi}_{AB},H_{A}^{\Gamma})={}_{AB}\!\bra{\psi}(H_{A}^{\Gamma})^{2}\otimes \mathbb{I}_{B}\ket{\psi}_{AB}-\big({}_{AB}\!\bra{\psi}H^\Gamma_{A}\otimes \mathbb{I}_{B}\ket{\psi}_{AB}\big)^{2}$. For pure states, the QFI reduces (up to a factor) to the variance, i.e.,~$\mathcal{F}(\ket{\psi}_{AB},H_{A}^{\Gamma})=4{\cal V}(\ket{\psi}_{AB},H_{A}^{\Gamma})$~\cite{paris2009quantum,toth2014quantum}, and so the IE is equal to the IP.

For a general mixed state $\rho_{AB}$, we use the standard convex-roof construction  to extend the definition of  the IE as
\begin{equation}\label{Eq:IEMixed}
\mathcal{E}^{\Gamma}(\rho_{AB})=\min_{\{p_{i},\ket{\psi_{i}}_{AB}\}}{\sum}_{i} p_{i} \mathcal{E}^{\Gamma}(\ket{\psi_{i}}_{AB}),
\end{equation}
considering all decompositions of $\rho_{AB}=\sum_{i}p_{i}\ket{\psi_{i}}_{AB}\!\bra{\psi_{i}}$  into pure states. We get that the IE is a full convex entanglement monotone, satisfying in particular the two key requirements stemming from the mathematical theory of entanglement as a resource \cite{horodecki2009quantum,plenio2007an}: (i) $\mathcal{E}^{\Gamma}(\rho_{AB})=0$ for all separable states of Eq.~(\ref{Eq:Sep}), and (ii) $\mathcal{E}^{\Gamma}(\rho_{AB}) \geq \sum_{i} q_{i} \mathcal{E}^{\Gamma}(\rho_{AB}^{(i)})$, with $q_{i} = \mbox{Tr}(K_{i}\rho_{AB} K_{i}^{\dagger})$ and $\rho_{AB}^{(i)} = K_{i}\rho_{AB} K_{i}^{\dagger}/q_{i}$, meaning that entanglement can never be generated or increased on average through local operations and classical communication (LOCC), where the product Kraus operators $\{K_i\}$ describe the action of a LOCC map, $\Lambda_{LOCC} (\rho_{AB}) = \sum_{i}K_{i} \rho_{AB} K_{i}^{\dagger}$. This holds by virtue of the convex-roof extension \cite{vidal2000entanglement}, given that the quantity in Eq.~(\ref{Eq:IEPure}) is a LOCC monotone for pure states \cite{girolami2013characterizing}. Together with convexity, the properties above imply
standard LOCC monotonicity for the IE, $\mathcal{E}^{\Gamma}(\rho_{AB}) \geq \mathcal{E}^{\Gamma}(\Lambda_{LOCC}(\rho_{AB}))$~\cite{vedral1998entanglement}.

\section{Hierarchy of interferometric figures of merit}\label{Sec:HI}
The IE can be understood to quantify the worst-case sensitivity from the family of generating Hamiltonians $H_{A}^{\Gamma}$ if one were to perform interferometry using individual pure probe states $\ket{\psi_{i}}_{AB}$ and then average the results with probabilities $p_{i}$. Conversely, the IP, as given by Eq.~(\ref{Eq:IP}), represents the worst-case sensitivity by using the mixed state $\rho_{AB}=\sum_i p_i \ket{\psi_i}_{AB}\bra{\psi_i}$ as a probe state directly.
Furthermore, by virtue of the convex-roof construction \cite{uhlmann2010entropy}, the extension of the IE to mixed states in Eq.~(\ref{Eq:IEMixed}) amounts to the largest convex function which reduces to the worst-case sensitivity (that is, to the IP) for pure states. This indicates that both quantifiers are defined and physically motivated within the same operational setting, which makes their comparison  meaningful.

Intuitively, one may then expect that the interferometric resource quantified by the IE can never exceed the figure of merit given by the IP due to the extra minimization over all pure-state decompositions of $\rho_{AB}$.  We will now see that this intuition is true \footnote{Note however that such a hierarchy is non-trivial and does not hold generally for a pair of quantifiers of entanglement and quantum correlations beyond entanglement, even if defined through similar methods as the two quantities considered here. For instance, in quantum information theory, two of the most common measures of entanglement and general quantum correlations are the entanglement of formation $E^f(\rho_{AB})$ and the quantum discord $D_A(\rho_{AB})$, respectively \cite{plenio2007an,ollivier2001quantum,adesso2016measures}. These two measures coincide on pure states (reducing to the entropy of entanglement), while the entanglement of formation is extended to mixed states via the convex-roof construction. However, there is no strict inequality between them, $E^f(\rho_{AB}) \lesseqqgtr D_A(\rho_{AB})$, meaning that one quantity may be larger or smaller than the other depending on the state $\rho_{AB}$ \cite{modi2012classical}. On the other hand, Eq.~(\ref{Eq:PreIPTangleInequality}) shows that interferometrically inspired measures of entanglement and quantum correlations beyond entanglement do obey a strict inequality, $\mathcal{E}^{\Gamma}(\rho_{AB}) \leq \mathcal{P}_{A}^{\Gamma}(\rho_{AB})$, for any state $\rho_{AB}$.}.
The first ingredient to use is that the QFI is (four times) the convex roof of the variance~\cite{toth2013extremal,yu2013quantum},
\begin{equation}
\mathcal{F}(\rho_{AB},H_{A}^{\Gamma}) = 4 \min_{\{p_{i},\ket{\psi_{i}}\}} {\sum}_{i}p_{i} {\cal V}(\ket{\psi_{i}}_{AB},H_{A}^{\Gamma}).
\end{equation}
Using this fact, and the definition of the IE, we can say that
\begin{eqnarray}
\mathcal{F}(\rho_{AB},H_{A}^{\Gamma}) &=& 4\min_{\{p_{i},\ket{\psi_{i}}\}} {\sum}_{i}p_{i} {\cal V}(\ket{\psi_{i}}_{AB},H_{A}^{\Gamma}) \nonumber \\
&\geq& 4\min_{\{p_{i},\ket{\psi_{i}}\}} {\sum}_{i}p_{i} \min_{H_{A}^{\Gamma}}{\cal V}(\ket{\psi_{i}}_{AB},H_{A}^{\Gamma}) \nonumber \\
&=& 4\min_{\{p_{i},\ket{\psi_{i}}\}} {\sum}_{i}p_{i} \mathcal{E}^{\Gamma}(\ket{\psi_{i}}_{AB}) \nonumber \\
&=& 4\,\mathcal{E}^{\Gamma}(\rho_{AB}),
\end{eqnarray}
where in the inequality we minimize over $H_{A}^{\Gamma}$, and in the second and third equalities we use the definition of IE in Eqs.~(\ref{Eq:IEPure}) and (\ref{Eq:IEMixed}). Finally, using the definition of IP in Eq.~(\ref{Eq:IP}), we have
\begin{equation}\label{Eq:PreIPTangleInequality}
\mathcal{P}_{A}^{\Gamma}(\rho_{AB})= \mbox{$\frac{1}{4}$}\min_{H_{A}^{\Gamma}} \mathcal{F}(\rho_{AB},H_{A}^{\Gamma}) \geq \min_{H_{A}^{\Gamma}} \mathcal{E}^{\Gamma}(\rho_{AB}) = \mathcal{E}^{\Gamma}(\rho_{AB}).
\end{equation}

We thus construct the fundamental hierarchy of resources:
\begin{equation}\label{Eq:IPTangleInequality}
\mbox{$\frac14$}\mathcal{F}(\rho_{AB},H_{A}^{\Gamma}) \geq \mathcal{P}_{A}^{\Gamma}(\rho_{AB}) \geq \mathcal{E}^{\Gamma}(\rho_{AB}), \quad \forall \, \rho_{AB}, H_A^{\Gamma},
\end{equation}
where the leftmost inequality holds by construction \cite{girolami2014quantum}, while the rightmost inequality is our first main result. This shows that entanglement (quantified by IE) does not entirely capture the figure of merit in quantum interferometry, as it accounts only for a portion of the relevant quantum correlations (quantified by IP), which in turn provide a tighter lower bound to the QFI for any $H_A^\Gamma$. Note that any pure state saturates the rightmost inequality. Equation~(\ref{Eq:IPTangleInequality}) also succeeds in unifying different nonclassical signatures under the operational umbrella of interferometric phase estimation.


\section{Interferometric power as a resource}\label{Sec:IP}
In the following, we investigate the validation of the IP as a measure of quantum correlations beyond entanglement. In \cite{girolami2014quantum}, the IP has been shown to obey the following properties: (i) $\mathcal{P}_{A}^{\Gamma}(\rho_{AB}) = 0$ iff $\rho_{AB}$ is a classical state of the form  Eq.~(\ref{Eq:CQ}), (ii) $\mathcal{P}_{A}^{\Gamma}(\rho_{AB})$ is invariant under local unitaries, (iii) $\mathcal{P}_{A}^{\Gamma}(\rho_{AB})$ reduces to an entanglement monotone for pure states (here identified as the IE), and (iv) $\mathcal{P}_{A}^{\Gamma}(\rho_{AB})$ is nonincreasing under the action of any local operation on subsystem $B$. These properties can be recognized as a set of necessary requirements for any good quantifier of quantum correlations \cite{streltsov2011linking,aaronson2013comparative,girolami2013characterizing,adesso2016measures}. However, adopting a resource-theory perspective \cite{horodecki2013quantumness,coecke2016mathematical,brandao2015second}, one should impose a more general monotonicity requirement,  that any measure of our resource should not increase when a suitable set of free operations is applied to any state.

 While LOCC are well established as the free operations for entanglement theory \cite{horodecki2009quantum,plenio2007an}, the corresponding set of free operations for more general quantum correlations has remained elusive. Recent findings have identified local commutativity-preserving operations (LCPO) as the maximal set of local operations unable to create quantum correlations from an initial classical state~\cite{streltsov2011behavior,hu2012necessary,guo2013necessary}. The LCPO $\Lambda_{LCPO} = \Lambda_{A} \otimes \mathbb{I}_{B}$ preserve commutativity of local states on $A$, i.e.,
$
[\Lambda_{A}(\rho_{A}),\Lambda_{A}(\varsigma_{A})] = 0 \,\,\,\, \forall \,\, \rho_{A}, \varsigma_{A} \, \mbox{such that} \,\, [\rho_{A},\varsigma_{A}] = 0$~\cite{footNote1}.
Monotonicity with respect to these operations has been proposed as an additional requirement for any measure of quantum correlations~\cite{adesso2016measures}, so it is crucial to establish whether the IP has this property, i.e.,~if $ \mathcal{P}_{A}^{\Gamma}(\rho_{AB}) \geq \mathcal{P}_{A}^{\Gamma}(\Lambda_{LCPO}(\rho_{AB}))$. We now prove that this is true when probe $A$ is a qubit.

Restricted to qubits on $A$, the commutativity-preserving operations are of two types, completely decohering and unital. The completely decohering operations map any state to one diagonal in a fixed reference basis $\{\ket{i}_{A}\}$, $\Lambda_{A}(\rho_{A}) = {\sum}_{i} p_{i}(\rho_{A}) \ket{i}\bra{i}_{A}$,
with $p_{i}(\rho_{A})$ probabilities dependent on $\rho_{A}$, while unital operations preserve the identity, $\Lambda_{A}(\mathbb{I}_{A}) = \mathbb{I}_{A}$. Observing that local completely decohering operations $\Lambda_{A}$ on $A$ always return a classical state, i.e.~$\Lambda_{A}(\rho_{AB})$ is of the form of Eq.~\ref{Eq:CQ} for any input $\rho_{AB}$, it trivially follows that $\mathcal{P}_{A}^{\Gamma}(\rho_{AB}) \geq \mathcal{P}_{A}^{\Gamma}(\Lambda_{A}(\rho_{AB})) = 0$ in this case. We then need to show monotonicity of the IP under LCPO with unital operations $\Lambda_{A}$ on $A$ to guarantee overall monotonicity for qubit-qudit probes. The proof is provided in Appendix~\ref{App:Monoqubit}.

When probe $A$ has dimension $d_{A} > 2$, the commutativity-preserving operations can be completely decohering (as above), or isotropic,
$\Lambda_{A}(\rho_{A}) = t \Phi(\rho_{A}) + (1-t) \mathbb{I}_{A}/d_{A}$,
where $\Phi(\rho_{A})$ is either a unitary operation, i.e.,~$\Phi(\rho_{A})=U_{A}\rho_{A}U_{A}^{\dagger}$ for some unitary $U_{A}$, or an antiunitary operation, i.e.~$\Phi(\rho_{A})=U_{A}\rho_{A}^{T}U_{A}^{\dagger}$, with $\rho_{A}^{T}$ denoting the transpose of $\rho_{A}$. For $\Lambda_{A}(\rho_{A})$ to be completely positive, $t$ is constrained to $t \in [\frac{-1}{d_A^{2}-1},1]$ when $\Phi$ is unitary, and $t \in [\frac{-1}{d_A-1},\frac{1}{d_A+1}]$ when $\Phi$ is antiunitary. We provide the operator-sum representation of $\Lambda_{A}(\rho_{A})$ in Appendix~\ref{App:Kraus}. Since again the completely decohering operations on $A$ always return a classical state, investigating monotonicity of the IP under LCPO when $A$ is a qudit requires testing monotonicity under isotropic operations $\Lambda_{A}$ on $A$. In Appendix~\ref{App:Monoiso}, we prove such monotonicity when $\Phi$ is unitary and $t \in [0,1]$, while the remaining cases are presently left as an open question.
These results show that the IP is a full quantum correlations monotone for arbitrary qubit-qudit states, and a valid monotone under a subset of LCPO for general qudit-qudit states.

\section{Experimental investigation of extremal states}\label{Sec:EX}
We finally explore the interplay between IP and IE as captured by Eq.~(\ref{Eq:IPTangleInequality}) via an in-depth numerical analysis supplemented by an experimental two-qubit NMR implementation using a BRUKER Ascend $600$-MHz spectrometer at room temperature.

\begin{figure}[b]
  \centering
    \includegraphics[width=8cm]{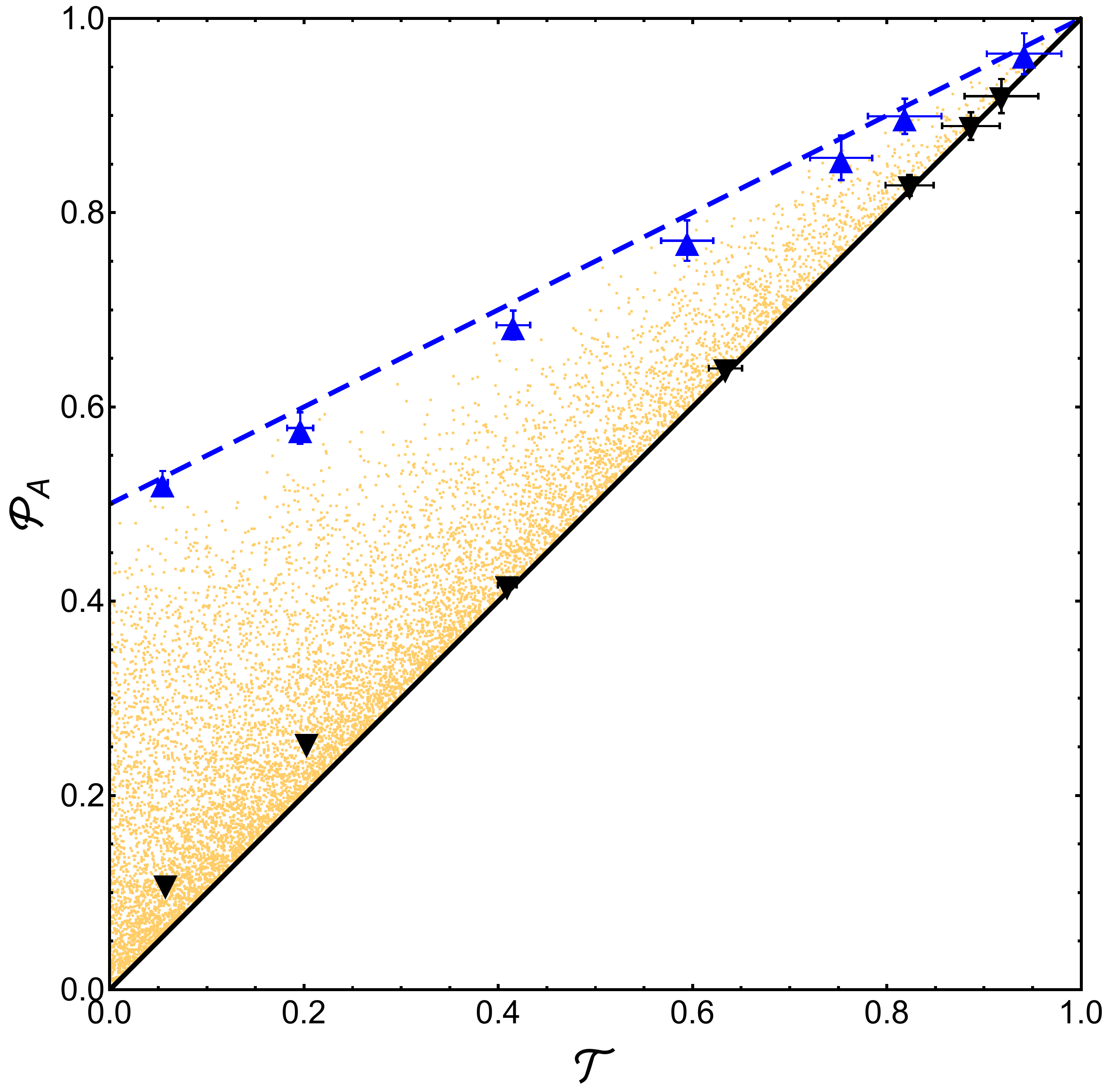}
 \caption{Comparison of the IP $\mathcal{P}_{A}(\rho_{AB})$ versus the IE, alias tangle $\mathcal{T}(\rho_{AB})$, for two-qubit rank-$2$ states $\rho_{AB}$. The lines correspond to the family of rank-$2$ $X$ states of Eq.~(\ref{Eq:FamilyOfStates}) parameterized by $\theta_{1}$ and $\theta_{2}$ for two cases:  $\mathcal{P}_{A}(\rho_{AB})=\mathcal{T} (\rho_{AB})$ (solid line), which is the smallest IP for a given tangle; and  $\mathcal{P}_{A}(\rho_{AB}) = \frac12[1+\mathcal{T} (\rho_{AB})]$ (dashed line), which is identified numerically as the largest IP for a given tangle amongst rank-2 states. The little squares depict $10^{5}$ randomly generated rank-$2$ states, which are always found within the region given by $\mathcal{T} (\rho_{AB}) \leq \mathcal{P}_{A}(\rho_{AB}) \leq \frac12[1+\mathcal{T} (\rho_{AB})]$. The triangles correspond to experimental two-qubit states of the form of Eq.~(\ref{Eq:FamilyOfStates}) prepared using an NMR setup, with angles $\theta_{1}$ and $\theta_{2}$ given in Table~\ref{Table:ThetaValues}, to approach the lower extremal boundary (downward triangles) and the upper extremal one (upward triangles). Errors bars are calculated as detailed in the text.
}
 \label{Figure:IPVsTangle}
\end{figure}

For two-qubit probes and a standard equispaced spectrum  $\Gamma = \{-1,1\}$ of our generating Hamiltonians $H_{A}^{\Gamma}$ (we will drop the superscript $\Gamma$ in what follows), the IE  reduces to the tangle $\mathcal{T}$ or squared concurrence \cite{hill1997entanglement,wootters1998entanglement,coffman2000distributed,rungta2001universal,rungta2003concurrence}, which is a monotonic function of the entanglement of formation (see Appendix~\ref{App:Equi}),
while the IP $\mathcal{P}_A$ also adopts a simple closed form~\cite{girolami2014quantum}. Focusing on probes $\rho_{AB}$ with rank-$2$ density matrices, we locate a family of states with the largest and smallest IP for a given tangle. This family is parameterized by two angles $\theta_{1}, \theta_{2} \in [0,\pi/2]$, and can be expressed in the computational basis as an $X$ state,
\begin{equation}
\!\! \rho_{AB} = \frac{1}{2}\left(
\begin{array}{cccc}
 c_+(\theta_1,\theta_2) &  0 &   0 &   d_+(\theta_1,\theta_2)  \\
 0 &   s_+(\theta_1,\theta_2) &   d_-(\theta_1,\theta_2) &   0 \\
  0 &   d_-(\theta_1,\theta_2) &   s_-(\theta_1,\theta_2) &   0 \\
  d_+(\theta_1,\theta_2) &   0 &   0 &   c_-(\theta_1,\theta_2)\end{array}\right),\!
\label{Eq:FamilyOfStates}
\end{equation}
with $c_\pm(\theta_1,\theta_2) = \cos^2\left(\frac{\theta_2}{2}\right)[1 \pm \sin(\theta_1)]$, $s_\pm(\theta_1,\theta_2) = \sin^2\left(\frac{\theta_2}{2}\right)[1 \pm \sin(\theta_1)] $, and $d_\pm(\theta_1,\theta_2) = -\frac{\cos(\theta_1)}{2}[1 \pm \cos(\theta_2)]$.

\begin{table}[b]
\begin{center}
\caption{Values of $\theta_{1}$ and $\theta_{2}$, applied with $x$ phase, for experimentally generated two-qubit states of Eq.~(\ref{Eq:FamilyOfStates}), approximating (a) lower extremal states with IP equal to tangle ($\theta_1=0$) and (b) upper extremal states with the largest IP for a given tangle among rank-$2$ states.}
\begin{tabular}{ccccccccc}
\hline \hline
{{(a)}} & \,\, $\theta_2$ \,\,\,\, & \,\, $\frac{\pi}{14}$ \,\,\,\,&\,\,\,\, $\frac{\pi}{7}$ \,\,\,\,&\,\,\,\, $\frac{3\pi}{14}$ \,\,\,\,&\,\,\,\, $\frac{2\pi}{7}$ \,\,\,\,&\,\,\,\, $\frac{5\pi}{14}$ \,\,\,\,&\,\,\,\, $\frac{3\pi}{7}$ \,\,\,\,&\,\,\,\, $\frac{\pi}{2}$ \,\,\,\, \\[3pt] \hline \multirow{ 2}{*}{{{(b)}}} &   $\theta_1$ &  $0$ & $\frac{7\pi}{90}$ & $\frac{\pi}{9}$ & $\frac{13\pi}{90}$ & $\frac{8\pi}{45}$ & $\frac{19\pi}{90}$ & $\frac{11\pi}{45}$ \\[3pt]
& $\theta_2$ & $0$ & $\frac{\pi}{10}$ & $\frac{\pi}{10}$ & $\frac{7\pi}{45}$ & $\frac{19\pi}{90}$ & $\frac{3\pi}{10}$ & $\frac{7\pi}{18}$ \\[3pt] \hline  \hline
\end{tabular}
\label{Table:ThetaValues}
\end{center}
\end{table}

One may calculate the tangle and IP of $\rho_{AB}$ to be
\begin{eqnarray}
\mathcal{T} (\rho_{AB}) &=& \cos^{2} (\theta_{1}) \cos^{2}(\theta_{2}),  \\
\mathcal{P}_{A}(\rho_{AB}) &=& \min \Bigg\{\! \cos (\theta_{1})^{2},\, \frac{3-\cos (2\theta_{1})+2 \cos^2 (\theta_{1})\cos (2 \theta_{2})}{4}\! \Bigg\} \nonumber .
\end{eqnarray}
In particular, whenever $\theta_{1} = 0$, we have that $\mathcal{P}_{A}(\rho_{AB})=\mathcal{T} (\rho_{AB})$, identifying an extremal subset of rank-$2$ states for which the rightmost inequality in Eq.~(\ref{Eq:IPTangleInequality}) is saturated (recall that this is also true for all pure states, i.e., rank-$1$ states). Furthermore, when $\theta_{1} = \arccos \left( \sqrt{\frac{1+T}{2}}\right)$ and $\theta_{2} = \frac{1}{2}\arccos\left( 3-\frac{4}{1+T} \right)$ for some $T \in [0,1]$, it holds that $\mathcal{T} (\rho_{AB})=T$ and $\mathcal{P}_{A}(\rho_{AB}) = \frac12[{1+\mathcal{T} (\rho_{AB})}]$, which we conjecture to be the maximum IP for a given tangle that can be reached among all rank-2 states \cite{noteX}. Figure~\ref{Figure:IPVsTangle} plots the IP and tangle for these two cases. The conjecture is further supported by the numerical investigation of  $10^{5}$ rank-$2$ states randomly drawn from the uniform distribution according to the Hilbert-Schmidt measure~\cite{bengtsson2007geometry}, whose corresponding points in the IP vs  tangle plane always lie in the triangular region between the two extremal cases (see Fig.~\ref{Figure:IPVsTangle}).

\begin{figure}[!t]
\begin{center}
\includegraphics[width=8.5cm]{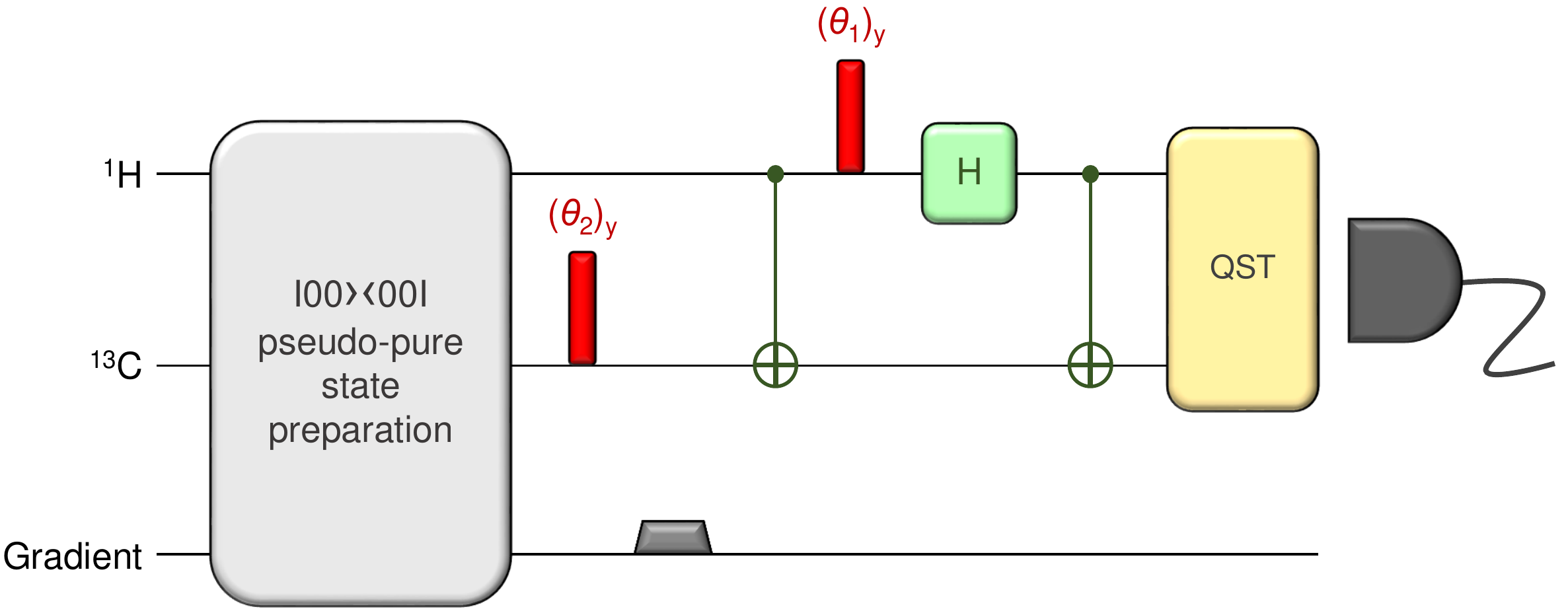}
\caption{\label{Figure:IPTseq} NMR pulse sequence to prepare two-qubit states of the form of Eq.~(\ref{Eq:FamilyOfStates}) using qubits encoded in $^1$H and $^{13}$C nuclear spins of chloroform. Following state preparation as described in the text, we perform four-pulse quantum state tomography (QST)~\cite{long2001analysis}.}
\end{center}
\end{figure}

In our experiment, the two-qubit system was encoded on $^1$H and $^{13}$C spin-$1/2$ nuclei in a chloroform (CHCl$_3$) enriched with a $^{13}$C sample, allowing complete  control of the amplitude and phase of each qubit separately \cite{isa1,isa2,isa3}.  Applying the pseudopure-state technique \cite{nielsen2010quantum, gershenfeld1997bulk, cory1997ensemble, knill1998effective, oliveira2011nmr}, the family of states in Eq.~(\ref{Eq:FamilyOfStates}) can be implemented easily in our NMR setup by transformations in the deviation matrix of the thermal configuration. These states are obtained from the NMR radiofrequency pulse sequence described in Fig.~\ref{Figure:IPTseq}, where the pseudopure state $\ket{00}\bra{00}$ is prepared as described in Ref.~\cite{oliveira2011nmr} (see Appendix~\ref{App:Exp}) and the quantum operations are implemented by controlled not (\textsc{cnot}) $[\frac{\pi}{2}]_y^{C}\rightarrow U[\frac{1}{2J}] \rightarrow [\frac{\pi}{2}]_x^{C}  \rightarrow [\frac{\pi}{2}]_{-y}^{C}  \rightarrow [\frac{\pi}{2}]_x^{C}  \rightarrow [\frac{\pi}{2}]_y^{C}  \rightarrow [\frac{\pi}{2}]_{-y}^{H}  \rightarrow [\frac{\pi}{2}]_{-x}^{H}  \rightarrow [\frac{\pi}{2}]_y^{H}$ and Hadamard-$[\frac{\pi}{2}]_y^{H}\rightarrow [\pi]_x^{H}$ \cite{oliveira2011nmr}. In particular, states approximating the two extremal rank-$2$ classes with maximum and minimum IP for a given tangle were prepared by varying the angles $\theta_1$ and $\theta_2$ according to Table~\ref{Table:ThetaValues}. After state preparation, we performed full four-pulse quantum state tomography: $\mathbb{I}^C\mathbb{I}^H,[\frac{\pi}{2}]_x^{C},[\frac{\pi}{2}]_y^{C}, [\frac{\pi}{2}]_x^{H}[\frac{\pi}{2}]_x^{C}$, as described in~\cite{long2001analysis}. The resultant Uhlmann fidelity~\cite{uhlmann1976transition} with the corresponding target state was found to be always larger than $95\%$. Error bars were estimated by propagating the errors in pulse calibration (smaller than $3\%$ per pulse, as determined by pulse-width fitting) for the entire pulse sequence and producing statistics based on $100$ computationally simulated runs per each preparation. Further experimental details are available in Appendix~\ref{App:Exp}.

\section{Summary and Outlook} \label{Sec:SU}
We reported results of fundamental and practical impact on the quantification of resources for quantum enhanced interferometry, advancing along three main paths.

First, we defined the IE, an entanglement monotone inspired by the figure of merit in interferometry, and showed that it can never exceed the IP, a quantifier of general quantum correlations introduced in \cite{girolami2014quantum}; this establishes a hierarchical relation between useful nonclassical resources, showing in particular the inability of entanglement to fully capture the precision available for estimating a phase $\varphi$ imprinted by a Hamiltonian with a fixed spectrum but variable eigenbasis. We remark that this hierarchy exists specifically between IE and IP (which can be rightfully compared because they are derived from similar principles) and does not necessarily extend to other pairs of measures of entanglement and quantum correlations beyond entanglement. A worthwhile development will be to extend this analysis to other phase-imprinting operations besides unitaries, such as noisy phase-covariant operations~\cite{smirne2016ultimate,nichols2016practical}, which are the free operations in a resource theory of quantum coherence viewed as asymmetry with respect to time translations~\cite{marvian2014extending,marvian2016quantum,streltsov2016colloquium}.

Second, the {\em bona fide} role of the IP in quantum information theory has been further cemented by showing its monotonicity (for all qubit-qudit states) under LCPO, a  postulated meaningful set of free operations for a resource theory of quantum correlations \cite{adesso2016measures}. Further developments may identify a different set of free operations, possibly motivated from additional physical restrictions. However, since LCPO form the maximal set of local operations unable to create quantum correlations \cite{streltsov2011behavior,hu2012necessary,guo2013necessary}, any such possible set of free operations must lie within LCPO, and monotonicity of the IP will remain. Our next steps will be to investigate the full monotonicity of the IP under LCPO when operating locally on probes with dimension higher than $2$, which will be the focus of future work.

Third, we investigated how far quantum correlations can go  {\it beyond} entanglement \cite{girolami2011interplay,piani2012quantumness} in two-qubit systems. We identified classes of extremal rank-$2$ states with maximum and minimum IP at given IE, and prepared instances of such states experimentally using a room-temperature NMR setup. This shows that bipartite probe states offering substantial extra gain in performance for interferometry given a fixed degree of entanglement are accessible in the laboratory. While maximum performance is always reached on pure maximally entangled states, finitely entangled states yet with extremal quantum correlations can be valuable whenever access to pure-state preparations is precluded. It will be interesting to further explore the practical usefulness of quantum correlations beyond entanglement in technological settings such as quantum interferometry, metrology, and discrimination \cite{bogaert2016metrological,adesso2016measures,braun2017quantum}, possibly with different experimental setups.

\vspace{20mm}\begin{acknowledgments}
This work was supported by the European Research Council (ERC) Starting Grant GQCOP (Grant No.~637352),  the Foundational Questions Institute's Physics of the Observer Programme (Grant No.~FQXi-RFP-1601), the Royal Society International Exchanges (Grant No.~IE150570) the CAPES Pesquisador Visitante Especial (Grant No.~108/2012), the CNPQ (Grants No.~312852/2014-2, No. 304955/2013-2, and No. 443828/2014-8), and the Brazilian National Institute of Science and Technology of Quantum Information (INCT/IQ). E.R.dA. thanks L. A. Colnago and the Brazilian Agricultural Research Corporation (EMBRAPA Instrumentation) for NMR machine time.
We thank K. Modi, G. T{\'o}th, and S. Yu for helpful discussions.
\end{acknowledgments}

\appendix

\section{Monotonicity of the IP under local unital maps for qubit-qudit systems}\label{App:Monoqubit}

We will prove that $\mathcal{P}_{A}^{\Gamma}(\rho_{AB}) \geq \mathcal{P}_{A}^{\Gamma}(\Lambda_{A} \otimes \mathbb{I}_{B}(\rho_{AB}))$ for qubit-qudit  states $\rho_{AB}$ and unital operations $\Lambda_{A}$ acting on $A$, i.e., where $\Lambda_{A}$ preserves the identity. Consider the dilation $\tau_{ABC}$ of $\Lambda_{A} \otimes \mathbb{I}_{B}(\rho_{AB})$ into a larger space including an extra ancillary system $C$, such that $\mbox{Tr}_{C}[\tau_{ABC}] = \Lambda_{A} \otimes \mathbb{I}_{B}(\rho_{AB})$~\cite{keyl2002fundamentals}. The following inequality holds:
\begin{equation}\label{Equation:DilationMonotonicity}
\mathcal{P}_{A}^{\Gamma}(\tau_{ABC}) \geq \mathcal{P}_{A}^{\Gamma}(\mbox{Tr}_{C}[\tau_{ABC}]) = \mathcal{P}_{A}^{\Gamma}(\Lambda_{A} \otimes \mathbb{I}_{B}(\rho_{AB}))\,,
\end{equation}
since the IP never increases under any operation on subsystems other than $A$~\cite{girolami2014quantum}. It is then sufficient to prove that $\mathcal{P}_{A}^{\Gamma}(\rho_{AB}) \geq \mathcal{P}_{A}^{\Gamma}(\tau_{ABC})$ to arrive at the desired inequality. To do this, we use the fact that any unital qubit operation can be equivalently written as a convex combination of unitaries (or random unitary channel)~\cite{mendl2009unital}, i.e.,
\begin{equation}
\Lambda_{A}(\rho_{A}) = \sum_{i} p_{i} U_{A}^{(i)} \rho_{A}( U_{A}^{(i)}) ^{\dagger}
\end{equation}
for some mixture of unitaries $\{U_{A}^{(i)}\}$ with probabilities $\{p_{i}\}$ acting on subsystem $A$ in the state $\rho_{A}$. This can be used to explicitly write the dilated state as
\begin{equation}
\tau_{ABC} = U_{ABC} (\rho_{AB} \otimes \ket{\alpha}\bra{\alpha}_{C}) U_{ABC}^{\dagger},
\end{equation}
with
\begin{eqnarray}
U_{ABC} &=& \sum_{i} U_{A}^{(i)} \otimes \mathbb{I}_{B} \otimes \ket{i}\bra{i}_{C}, \nonumber \\
\ket{\alpha}_{C} &=& \sum_{i} \sqrt{p_{i}} \ket{i}_{C}.
\end{eqnarray}

We now make use of the explicit form of the IP for qubit-qudit states given in~\cite{girolami2014quantum}, $\mathcal{P}_{A}^{\Gamma}(\rho_{AB}) = \alpha^{2} \min{\{\lambda_{i}\}}$, where $\{\lambda_{i}\}$ are the eigenvalues of the $3 \times 3$ matrix
\begin{equation}\label{Eq:IPClosed}
M = \frac{1}{2}  \sum_{m,n:q_{m}+q_{n} \neq 0} \frac{(q_{m}-q_{n})^{2}}{q_{m}+q_{n}} \braket{\phi_{m}|\vec{\sigma}_{A}\otimes \mathbb{I}_{B}|\phi_{n}}\braket{\phi_{n}|\vec{\sigma}_{A}^{T}\otimes \mathbb{I}_{B}|\phi_{m}},
\end{equation}
with $q_{m}$ and $\ket{\phi_{m}}_{AB}$ being the eigenvalues and normalized eigenvectors of $\rho_{AB}$ and $\vec{\sigma}_A$ being the vector of the three Pauli matrices. We write any two-level spectrum as $\Gamma = \{\beta-\alpha,\beta+\alpha\}$, with $\alpha,\beta\in\mathbb R$. For convenience, in the following we set $\alpha = 1$ and $\beta = 0$ and consider the standard equispaced spectrum $\{-1,1\}$, but the proof holds for any $\alpha$ and $\beta$.  
The task is then to calculate the matrix $M'$ corresponding to $\tau_{ABC}$. The eigenvalues of $\tau_{ABC}$ are the same as those of $\rho_{AB}$, while the eigenvectors are given by
\begin{equation}
\ket{\Phi_{m}}_{ABC}=U_{ABC} \ket{\phi_{m}}_{AB}\otimes \ket{\alpha}_{C} .
\end{equation}

\begin{widetext}
We can then write
\begin{eqnarray}
M' &=& \frac{1}{2}\sum_{mn} \frac{(q_{m}-q_{n})^{2}}{q_{m}+q_{n}} \braket{\Phi_{m}|\vec{\sigma}_{A}\otimes \mathbb{I}_{BC}|\Phi_{n}}\braket{\Phi_{n}|\vec{\sigma}_{A}^{T}\otimes \mathbb{I}_{BC}|\Phi_{m}} \nonumber \\
&=& \frac{1}{2}\sum_{mn} \frac{(q_{m}-q_{n})^{2}}{q_{m}+q_{n}} \bra{\phi_{m}}_{AB} \otimes \bra{\alpha}_{C} U_{ABC}^{\dagger}\vec{\sigma}_{A}\otimes \mathbb{I}_{BC}U_{ABC} \ket{\phi_{n}}_{AB}\otimes \ket{\alpha}_{C}     \bra{\phi_{n}}_{AB} \otimes \bra{\alpha}_{C} U_{ABC}^{\dagger}\vec{\sigma}_{A}^{T}\otimes \mathbb{I}_{BC}U_{ABC} \ket{\phi_{m}}_{AB}\otimes \ket{\alpha}_{C} \nonumber \\
&=& \frac{1}{2}\sum_{mn} \frac{(q_{m}-q_{n})^{2}}{q_{m}+q_{n}} \braket{\phi_{m}|\sum_{i}p_{i}(U_{A}^{(i)})^{\dagger}\vec{\sigma}_{A} U_{A}^{(i)} \otimes \mathbb{I}_{B}|\phi_{n}}\braket{\phi_{n}|\sum_{j}p_{j}(U_{A}^{(j)})^{\dagger}\vec{\sigma}_{A}^{T} U_{A}^{(j)} \otimes \mathbb{I}_{B}|\phi_{m}} ,
\end{eqnarray}
\end{widetext}
where we have used the fact that $U_{ABC}\ket{\alpha}_{C} = \sum_{i} \sqrt{p_{i}} U_{A}^{(i)} \otimes \mathbb{I}_{B} \ket{i}_{C}$. From the well known correspondence between the special unitary group ${\sf SU}(2)$ and special orthogonal group ${\sf SO}(3)$, we can see that for each $i$ there exists an orthogonal matrix $R_{i}$ such that $(U_{A}^{(i)})^{\dagger}\vec{\sigma}_{A} U_{A}^{(i)} = R_{i} \vec{\sigma}_{A}$. We thus obtain
\begin{equation}
M' = L M L^{T},
\end{equation}
where $L=\sum_{i} p_{i} R_{i}$ is a real matrix such that $L^{T}L \leq \mathbb{I}$. 

Finally, let us consider the eigenvalues of $M'$. If $L$ is noninvertible, we know that $M'$ has a zero eigenvalue, and hence $\mathcal{P}_{A}^{\Gamma}(\rho_{AB}) \geq \mathcal{P}_{A}^{\Gamma}(\tau_{ABC}) = 0$. Instead, if $M'$ is invertible, consider the unit vector $\ket{v}$ constructed by
\begin{equation}
\ket{v} = \frac{(L^{T})^{-1}\ket{v}_{0}}{{\left|\left|(L^{T})^{-1}\ket{v}_{0}\right|\right|}} ,
\end{equation}
where $\ket{v}_{0}$ is the eigenvector of $M$ corresponding to the smallest eigenvalue $\lambda_{\min} \equiv \min \{\lambda_{i}\}$ of $M$.
It is then simple to see that
\begin{equation}
\lambda_{\min}' \leq \braket{v|M'|v} = \frac{\lambda_{\min}}{\left|\left|(L^{T})^{-1}\ket{v}_{0}\right|\right|^2} \leq \lambda_{\min},
\end{equation}
where $\lambda_{\min}'$ is the minimum eigenvalue of $M'$ and we have used the fact that $\left|\left|(L^{T})^{-1}\ket{v}_{0}\right|\right|\geq 1$ since $L^{T}L \leq \mathbb{I}$. 
Combined with Eq.~(\ref{Equation:DilationMonotonicity}), we then have that
\begin{equation}
\mathcal{P}_{A}^{\Gamma}(\rho_{AB}) = \lambda_{\min} \geq \lambda_{\min}' = \mathcal{P}_{A}^{\Gamma}(\tau_{ABC}) \geq \mathcal{P}_{A}^{\Gamma}(\Lambda_{A} \otimes \mathbb{I}_{B}(\rho_{AB})),
\end{equation}
establishing the monotonicity of the IP under qubit unital operations on probe $A$.

\section{Operator-sum representation of isotropic operations}\label{App:Kraus}

Here we provide explicitly the Kraus decomposition of the isotropic operations~\cite{hu2012necessary}
\begin{equation}
\Lambda_{A}(\rho_{A}) = t \Phi(\rho_{A}) + (1-t) \frac{\mathbb{I}_{A}}{d},
\end{equation}
for the two cases of unitary $\Phi$ and antiunitary $\Phi$, with $d$ being the dimension of system $A$. In particular, we shall provide the Kraus decomposition when $U_{A} = \mathbb{I}_{A}$. To find the Kraus decomposition for a general $U_{A}$, one simply needs to transform the following Kraus operators by $K_{i} \rightarrow U_{A} K_{i}$. Note that quantum correlations are invariant under local unitary transformations $U_{A}$, and so for our purposes it is sufficient to treat only the case $U_{A} = \mathbb{I}_{A}$.

Let us denote by $\{K_{i}\}$ the Kraus operators for the operator-sum representation of $\Lambda_{A}$,
\begin{equation}
\Lambda_{A} (X) = \sum_{i} K_{i} X K_{i}^{\dagger},
\end{equation}
where the condition $\sum_{i} K_{i}^{\dagger} K_{i} = \mathbb{I}_{A}$ must be satisfied for $\Lambda_{A}$ to be completely positive and trace preserving. In the following we shall first determine the allowed range of the parameter $t$ for both the unitary and antiunitary cases by imposing positivity of the Choi state. We introduce an ancilla $A'$ which is a copy of $A$; for brevity we shall indicate the computational basis of the joint $AA'$ system as $\ket{k,l}\equiv\ket{k}_A\otimes\ket{l}_{A'}$. The Choi state is then given by
\begin{equation}
\tau=\Lambda_A\otimes\mathcal{I}_{A'}(\ket{\Psi_+}\bra{\Psi_+}),
\end{equation}
where $\ket{\Psi_+}=\frac{1}{\sqrt{d}}\sum_{k=0}^{d-1}\ket{k,k}$ and $\mathcal I$ indicates the identity superoperator. Having imposed $\tau\ge0$, we will
then report the Kraus operators of $\Lambda_A$ and verify their completeness.

\subsection{Unitary case}
First we consider the unitary $\Phi$ case. We only need concern ourselves only with maps featuring $U_A=\mathbb I_A$. For any bipartite state $\rho_{AA'}$ we thus have $\Lambda_A\otimes\mathcal{I}_{A'}(\rho_{AA'})=t\rho_{AA'}+(1-t)\frac{\mathbb I_A}{d}\otimes {\sf Tr}_A(\rho_{AA'})$. The corresponding Choi state reads
\begin{equation}
\tau=t\ket{\Psi_+}\bra{\Psi_+}+\frac{1-t}{d^2}\mathbb I_A\otimes\mathbb I_{A'}.
\end{equation}
From the above we easily conclude that the spectrum of $\tau$ is $\{t+(1-t)/d^2, (1-t)/d^2\}$. Requiring the latter to be non-negative, we obtain the allowed range
\begin{equation}
-\frac{1}{d^2-1}\le t\le 1,\label{unitaryrange}
\end{equation}
which is tighter than what was reported in~\cite{hu2012necessary}. We shall now provide an explicit Kraus representation of the map. Consider the $d^{2}-1$ generalized Pauli matrices $\{\gamma_{i}\}_{i=1}^{d^{2}-1}$~\cite{bertlmann2008bloch}, and fix the $d$-dimensional identity matrix as $\gamma_{0} = \mathbb{I}_{A}$. The $d^{2}$ Kraus operators $\{K_{i}\}_{i=0}^{d^{2}-1}$ are then
\begin{eqnarray}
K_{0} &=& \sqrt{\frac{1+(d^{2}-1)t}{d^{2}}}\gamma_{0}, \nonumber \\
K_{i} &=& \sqrt{\frac{1-t}{2d}}  \gamma_{i} \,\,\,\,\,\,\,\,\,\,\,\,\,\,\,\,\,\,\,\,\,\,\, \forall i \in \{1,2,\ldots d^{2} -1\}.
\end{eqnarray}

We can now verify the condition $\sum_{i=0}^{d^{2}-1} K_{i}^{\dagger} K_{i} = \mathbb{I}_{A}$. Since the Kraus operators are Hermitian, and since $\gamma_{0}^{2} = \mathbb{I}_{A}$ and $\sum_{i=1}^{d^{2}-1} \gamma_{i}^{2}= \frac{2(d^{2}-1)}{d}\mathbb{I}_{A}$, we have
\begin{eqnarray}
\!\!\!\!\!\!\!\!\!\!\sum_{i=0}^{d^{2}-1} K_{i}^{\dagger} K_{i} &=& \left( \left|\frac{1+(d^{2}-1)t}{d^{2}}\right| + \frac{2(d^{2}-1)}{d} \times \left|\frac{1-t}{2d} \right| \right) \mathbb{I}_{A} \nonumber \\
\!\!\!\!\!\!\!\!\!\!&=& \frac{1}{d^{2}}\left( \left|1+(d^{2}-1)t\right| + (d^{2}-1) \left| (1-t)  \right| \right) \mathbb{I}_{A}.
\end{eqnarray}
Exploiting Eq.~\eqref{unitaryrange} we may simplify $\left|1+(d^{2}-1)t\right| = 1+(d^{2}-1)t$
and $\left| 1-t  \right|=1-t$, hence
\begin{eqnarray}
\sum_{i=0}^{d^{2}-1} K_{i}^{\dagger} K_{i} &=& \frac{1}{d^{2}}\left[ 1+(d^{2}-1)t + (d^{2}-1) (1-t) \right] \mathbb{I}_{A} \nonumber \\
&=& \mathbb{I}_{A}.
\end{eqnarray}

\subsection{Antiunitary case}
Now we treat the more complicated case of $\Phi$ being antiunitary (again, fixing $U_A=\mathbb I_A$). We thus have $\Lambda_A\otimes\mathcal{I}_{A'}(\rho_{AA'})=t\rho_{AA'}^{T_A}+(1-t)\frac{\mathbb I_A}{d}\otimes {\sf Tr}_A(\rho_{AA'})$, where $T_A$ indicates partial transposition on system $A$. The corresponding Choi state reads
\begin{equation}
\tau=\frac{t}{d}\sum_{k,l=0}^{d-1}\ket{k,l}\bra{l,k}+\frac{1-t}{d^2}\mathbb I_A\otimes\mathbb I_{A'}.
\end{equation}
By inspection we find that the eigenvectors of $\tau$ are in this case $\ket{k,k}$, with $k=0,...,d-1$, and $\frac{1}{\sqrt 2}(\ket{k,l}\pm\ket{l,k})$ for all pairs $k<l$. The spectrum of $\tau$ is then $\{(1-t)/d^2,t/d+(1-t)/d^2, -t/d+(1-t)/d^2\},$ from which we derive the constraint
\begin{equation}
-\frac{1}{d-1}\le t\le \frac{1}{d+1}.\label{anticase}
\end{equation}
As before, to write down a Kraus decomposition we can use the generalized Pauli matrices $\{\gamma_{i}\}_{i=1}^{d^{2}-1}$ with the identity $\gamma_{0} = \mathbb{I}_{A}$. Now, consider the set of vectorizations of the generalized Pauli matrices, $\{\vec{v}_{i}\}_{i=1}^{d^{2}-1}$ with $\vec{v}_{i} = \mbox{vec}(\gamma_{i})$, where $\mbox{vec}(X)=(\braket{0|X|0},\braket{0|X|1},\ldots, \braket{0|X|d}, \braket{1|X|0}, \braket{1|X|1},\ldots , \braket{d|X|d})$ is the vectorization of a matrix. We can split the generalized Pauli matrices into two categories based on their corresponding vectorizations: (1) $\mbox{sgn}(\vec{v}_{i} .\vec{v}_{i})=1$ and (2) $\mbox{sgn}(\vec{v}_{i} .\vec{v}_{i})=-1$. There are $(d+2)(d-1)/2$ generalized Pauli matrices of type 1 and $d(d-1)/2$ of type 2, and we call the generalized Pauli matrices of type 1: $\{\gamma_{i}^{(1)}\}_{i=1}^{(d+2)(d-1)/2}$ and those of type 2: $\{\gamma_{i}^{(2)}\}_{i=1}^{d(d-1)/2}$. Now we can give the Kraus decomposition:
\begin{eqnarray}
K_{0} &=& \sqrt{\frac{1+(d-1)t}{d^{2}}}\gamma_{0},  \\
K_{i} &=& \sqrt{\frac{1+(d-1)t}{2d}}  \gamma_{i}^{(1)} \quad \forall i \in \mbox{$\{1,2,\ldots , \frac{(d+2)(d-1)}{2}\}$}, \nonumber \\
K_{i+\frac{(d+2)(d-1)}{2}} &=& \sqrt{\frac{1-(d+1)t}{2d}}  \gamma_{i}^{(2)} \quad  \forall i \in \mbox{$\{1,2,\ldots , \frac{d(d-1)}{2}\}$}. \nonumber
\end{eqnarray}

We can also consider the condition $\sum_{i=0}^{d^{2}-1} K_{i}^{\dagger} K_{i} = \mathbb{I}_{A}$. Since the Kraus operators are Hermitian, and since $\gamma_{0}^{2} = \mathbb{I}_{A}$, $\sum_{i=1}^{(d+2)(d-1)/2} (\gamma_{i}^{(1)})^{2}= \frac{d^{2}+d-2}{d}\mathbb{I}_{A}$, and $\sum_{i=1}^{d(d-1)/2} (\gamma_{i}^{(2)})^{2}= (d-1)\mathbb{I}_{A}$, we have
\begin{eqnarray*}
\sum_{i=0}^{d^{2}-1}\! K_{i}^{\dagger} K_{i} &=& \left( \left|\frac{1+(d-1)t}{d^{2}}\right| +  \frac{d^{2}+d-2}{d} \times \left| \frac{1+(d-1)t}{2d}\right|\right.\\
&&\left.  + (d-1) \left|\frac{1-(d+1)t}{2d}\right| \right)\mathbb{I}_{A} \nonumber \\
&=& \frac{1}{d^{2}}\left( \left|1+(d-1)t\right| +  \frac{(d^{2}+d-2)}{2} \left| 1+(d-1)t\right|  \right. \\
&& \left.  + \frac{d^{2}-d}{2} \left|1-(d+1)t\right| \right)\mathbb{I}_{A} \nonumber \\
&=& \frac{1}{2d^{2}}\! \left[ (d^{2}+d)\left|1+(d-1)t\right| \right. \nonumber \\ & &  \left.    + (d^{2}-d) \left|1-(d+1)t\right| \right].
\end{eqnarray*}
Then, we may use Eq.~\eqref{anticase} to simplify $\left|1+(d-1)t\right|=1+(d-1)t$ and $\left|1-(d+1)t\right|=1-(d+1)t$, yielding $\sum_{i=0}^{d^{2}-1} K_{i}^{\dagger} K_{i} = \mathbb{I}_{A}$.

\section{Monotonicity of the IP under unitary isotropic operations}\label{App:Monoiso}

We now will prove that $\mathcal{P}_{A}^{\Gamma}(\rho_{AB}) \geq \mathcal{P}_{A}^{\Gamma}(\Lambda_{A} \otimes \mathbb{I}_{B}(\rho_{AB}))$ when $A$ has dimension larger than $2$ and for isotropic operations with unitary $\Phi$ and $t \in [0,1]$. For this range of $t$, we have that $\Lambda_{A} \otimes \mathbb{I}_{B}(\rho_{AB})$ is just a convex combination between $U_{A} \otimes \mathbb{I}_{B} \rho_{AB} U_{A}^{\dagger} \otimes \mathbb{I}_{B}$ and $\mathbb{I}_{A}/d_{A} \otimes \mbox{Tr}_{A}(\rho_{AB})$. From the convexity of the QFI~\cite{yu2013quantum} it holds that
\begin{eqnarray}
&&\mathcal{F}\left( \Lambda_{A} \otimes \mathbb{I}_{B}(\rho_{AB}),H_{A}^{\Gamma}\otimes \mathbb{I}_{B}\right)\nonumber \\
&=&\mathcal{F}\left(t U_{A} \otimes \mathbb{I}_{B} \rho_{AB} U_{A}^{\dagger} \otimes \mathbb{I}_{B} + (1-t) \frac{\mathbb{I}_{A}}{d_{A}} \otimes \mbox{Tr}_{A}(\rho_{AB}),H_{A}^{\Gamma}\otimes \mathbb{I}_{B}\right) \nonumber \\
&\leq& t \mathcal{F} \left(U_{A} \otimes \mathbb{I}_{B} \rho_{AB} U_{A}^{\dagger} \otimes \mathbb{I}_{B}, H_{A}^{\Gamma}\otimes \mathbb{I}_{B}\right) \nonumber \\
&& + (1-t)\mathcal{F} \left(\frac{\mathbb{I}_{A}}{d_{A}} \otimes \mbox{Tr}_{A}(\rho_{AB}), H_{A}^{\Gamma}\otimes \mathbb{I}_{B}\right) \nonumber \\
&=&t \mathcal{F} \left(U_{A} \otimes \mathbb{I}_{B} \rho_{AB} U_{A}^{\dagger} \otimes \mathbb{I}_{B}, H_{A}^{\Gamma}\otimes \mathbb{I}_{B}\right) \nonumber \\
&\leq& \mathcal{F} \left(U_{A} \otimes \mathbb{I}_{B} \rho_{AB} U_{A}^{\dagger} \otimes \mathbb{I}_{B}, H_{A}^{\Gamma}\otimes \mathbb{I}_{B}\right),
\end{eqnarray}
where in the second equality we use the fact that $\mathcal{F} \left(\frac{\mathbb{I}_{A}}{d_{A}} \otimes \mbox{Tr}_{A}(\rho_{AB}), H_{A}^{\Gamma}\otimes \mathbb{I}_{B}\right)=0$, which follows by noting that $\left[\frac{\mathbb{I}_{A}}{d_{A}} \otimes \mbox{Tr}_{A}(\rho_{AB}), H_{A}^{\Gamma}\otimes \mathbb{I}_{B}\right]=0$. Using the above inequality, we arrive at the monotonicity of the IP,
\begin{eqnarray}
\mathcal{P}_{A}^{\Gamma}(\Lambda_{A} \otimes \mathbb{I}_{B}(\rho_{AB})) &=& \frac{1}{4} \min_{H_{A}^{\Gamma}} \mathcal{F}\left( \Lambda_{A} \otimes \mathbb{I}_{B}(\rho_{AB}),H_{A}^{\Gamma}\otimes \mathbb{I}_{B}\right) \nonumber \\
&\leq & \frac{1}{4} \min_{H_{A}^{\Gamma}} \mathcal{F} \left(U_{A} \otimes \mathbb{I}_{B} \rho_{AB} U_{A}^{\dagger} \otimes \mathbb{I}_{B}, H_{A}^{\Gamma}\otimes \mathbb{I}_{B}\right) \nonumber \\
&=& \mathcal{P}_{A}^{\Gamma}(U_{A} \otimes \mathbb{I}_{B} \rho_{AB} U_{A}^{\dagger} \otimes \mathbb{I}_{B}) \nonumber \\
&=& \mathcal{P}_{A}^{\Gamma}(\rho_{AB}),
\end{eqnarray}
where in the third equality we use the invariance of the IP under local unitary transformations.

\section{Equivalence between the IE and the $I$-tangle for qubit-qudit states}\label{App:Equi}

It is now shown that the IE of Eq.~(5) in the main text reduces to the $I$-tangle defined in~\cite{rungta2001universal,rungta2003concurrence} when one considers qubit-qudit states. In particular, for two-qubit states, the IE becomes the standard tangle (squared concurrence)~\cite{coffman2000distributed}.

Consider the IE for pure states $\ket{\psi}_{AB}$ and set $H_{A}^{\Gamma} = \vec{n} \cdot \vec{\sigma}$ for some unit vector $\vec{n}$ and the Pauli vector $\vec{\sigma}$, which is the most general way to write a qubit Hamiltonian with spectrum $\Gamma$ equal to $\{-1,1\}$. We then have that
\begin{eqnarray}
\mathcal{E}^{\Gamma}(\ket{\psi}_{AB})&=&\min_{H_{A}^{\Gamma}}{\cal V}(\ket{\psi_{AB}},H_{A}^{\Gamma}) \nonumber \\
&=& \min_{H_{A}^{\Gamma}} \left[ \braket{\psi_{AB}|(H_{A}^{\Gamma})^{2} \otimes \mathbb{I}_{B}|\psi_{AB}} - \braket{\psi_{AB}|H_{A}^{\Gamma} \otimes \mathbb{I}_{B}|\psi_{AB}}^{2}\right] \nonumber \\
&=& \min_{H_{A}^{\Gamma}} \left[ 1 - \braket{\psi_{AB}|H_{A}^{\Gamma} \otimes \mathbb{I}_{B}|\psi_{AB}}^{2}\right] \nonumber \\
&=& 1- \max_{H_{A}^{\Gamma}}\braket{\psi_{AB}|H_{A}^{\Gamma} \otimes \mathbb{I}_{B}|\psi_{AB}}^{2} \nonumber \\
&=& 1 - \max_{i} \mu_{i},
\end{eqnarray}
where in the third equality we use the fact that $(H_{A}^{\Gamma})^{2} = (\vec{n} \cdot \vec{\sigma}_A)^{2} = \mathbb{I}_{A}$ and in the fifth equality we set $\mu_{i}$ to be the eigenvalues of $\vec{r} \vec{r}^{T}$, with $\vec{r} = \braket{\psi_{AB}|\vec{\sigma}_A \otimes \mathbb{I}_{B}|\psi_{AB}}$ being the local Bloch vector of $\ket{\psi}_{AB}$ on $A$. For any vector $\vec{v}$ of unit norm, $\vec{v} \cdot \vec{r} \vec{r}^{T} \cdot \vec{v} =( \vec{v} \cdot \vec{r})^{2} \leq ||\vec{r}||^{2},$
where the equality can be saturated by choosing $\vec{v}$ parallel to $\vec r$. Hence $\mu_{\max} \equiv \max_{i} \mu_{i} = ||\vec{r}||^{2}$, so that $\mathcal{E}^{\Gamma}(\ket{\psi}_{AB}) = 1 - ||\vec{r}||^{2}$.

Furthermore, it can be shown that
\begin{equation}
\frac{1+||\vec{r}||^{2}}{2} = \mbox{Tr}(\rho_{A}^{2}),
\end{equation}
with $\rho_{A} = \mbox{Tr}_{B}(\ket{\psi}\bra{\psi}_{AB})$ being the local state of subsystem $A$. Overall, we then have
\begin{equation}
\mathcal{E}^{\Gamma}(\ket{\psi}_{AB}) = 2[ 1 - \mbox{Tr}(\rho_{A}^{2})],
\end{equation}
which is ($2$ times) the local linear entropy of $\ket{\psi}_{AB}$. The $I$-tangle of~\cite{rungta2001universal,rungta2003concurrence} is defined for pure states as $2$ times the local linear entropy and for mixed states via the convex-roof construction. Hence, it is clear that in the case of qubit-qudit systems with a fixed spectrum $\{-1,1\}$, the IE is identical to the $I$-tangle. For two-qubit systems, the $I$-tangle is equal to the standard tangle~\cite{rungta2003concurrence,coffman2000distributed}:
\begin{equation}
\mathcal{T} (\rho_{AB}) = \max \{0,\lambda_{1}-\lambda_{2}-\lambda_{3}-\lambda_{4}\}^{2},
\end{equation}
where $\{\lambda_{i}\}$ are the eigenvalues of $\sqrt{\sqrt{\rho_{AB}}\tilde{\rho}_{AB}\sqrt{\rho_{AB}}}$ in nonincreasing order, with $\tilde{\rho}_{AB}=(\sigma_{y} \otimes \sigma_{y})\rho_{AB}^{T} (\sigma_{y} \otimes \sigma_{y})$ and $\sigma_{y}$ being the second Pauli matrix.

\section{NMR experimental details}\label{App:Exp}
The two-qubit system was associated with the $^1$H and $^{13}$C nuclear spins contained in a carbon-$13$ enriched chloroform sample (CHCl$_3$). This sample was prepared by mixing $100$ mg of $99\%$ $^{13}$C-labeled CHCl$_3$ in $0.7$ mL of $99.8\%$ CDCl$_3$ (both compounds were provided by Cambridge Isotope Laboratories Inc.). The experiments were performed at room temperature (around $25^{\mathrm{o}}$C) in a BRUKER Ascend $600$-MHz spectrometer located at the Brazilian Agricultural Research Corporation (EMBRAPA Instrumentation, S\~ao Carlos, Brazil). The spectrometer was equipped with a $5$-mm double-resonance probe head with field gradient coils. In CHCl$_3$, $^1$H and $^{13}$C are subjected to a small scalar spin-spin coupling of $J \approx 215$ Hz.

The thermal configuration of a NMR system is given by the density operator, $\rho_{eq} = \frac{1}{4}(\mathbb{I}_{AB} + \epsilon\,\sigma_z\otimes\sigma_z)$, where $\epsilon = \hbar\omega_L/4k_BT \sim 10^{-5}$. The deviation matrix $\Delta\rho_{eq} = \frac{1}{4}\sigma_z\otimes\sigma_z$ is the term of interest, as all the unitary transformations affect only this part. To prepare the state described in Eq. (\ref{Eq:FamilyOfStates}), first a $\ket{00}\bra{00}$ pseudopure state was prepared applying the pulse sequence
$\rho_{00}-[\frac{\pi}{3}]_x^{C}\rightarrow G_z(\tau) \rightarrow [\frac{\pi}{4}]_x^{H} \rightarrow U[\frac{1}{2J}] \rightarrow [\frac{\pi}{4}]_{-y}^{H} \rightarrow G_z(\tau)$ to the thermal equilibrium state \cite{oliveira2011nmr}. Here $G_z(\tau)$ corresponds to a gradient pulse applied for enough time to eliminate off-diagonal terms of the density matrix, and $U(1/2J)$ represents a free evolution under $J$ coupling for a period of $1/2J$ seconds. This step is followed by a pulse sequence in which each combination of $\theta_1$ and $\theta_2$ (described in Table~\ref{Table:ThetaValues}) provides one of the experimental points in Fig.~\ref{Figure:IPVsTangle}. Hadamard and \textsc{cnot} gates are implemented as described in the main text, as is the quantum state tomography procedure.

The error bars were estimated simulating the state preparation considering that each pulse was affected by an aleatory error, which was evaluated by pulse width (smaller than $3\%$ for all pulses). The simulation was repeated $100$ times, and the error was given by the distance between the theoretical state and the mean value of the error-affected states.


\begin{thebibliography}{78}
\bibitem [{foo()}]{footNote1}%
  \BibitemOpen
  \href@noop {} {}\bibinfo {note} {Note that property (iv) ensures monotonicity
  under any local operation on $B$. Hence, when considering free operations for
  quantum correlations, we can focus only on local operations on
  $A$.}\BibitemShut {Stop}%
\bibitem [{not()}]{noteX}%
  \BibitemOpen
  \href@noop {} {}\bibinfo {note} {By using Lagrange multipliers, one can prove
  rigorously that such states maximize the IP at given tangle among all
  two-qubit $X$ states of rank $2$. However, instances of states with ranks $3$
  and $4$ can be found with even larger IP at fixed nonzero
  tangle.}\BibitemShut {Stop}%
\end{thebibliography}

%

\end{document}